# Modeling of through-reactors with allowance of Large-Scale Effect on Heat and Mass Efficiency of Chemical Apparatuses

A.M. Brener, L.M. Musabekova

**Introduction**

The first considered problem is the problem of arising dynamic dissipative structures in the chemical reactor volume. The mathematical model of heat and mass transfer in the tubular through-reactors has been submitted. The model consists of the diffusion-kinetic equations for two reagents under the first-order reversible reaction and heat transfer equation with allowance for the latent heat of reaction.

Reaction rate constants for direct and reverse stages are related to temperature according to the Arrhenius law. The cases of both the adiabatic and non-adiabatic reactors have been considered.

We consider the chemical conversion proceeding under the scheme of a reversible first-order reaction in a tubular through-reactor:

$$X \underset{k_2}{\overset{k_1}{\rightleftarrows}} Y , \qquad (1)$$

where $k_1$ and $k_2$ are the reaction rate constants for direct and reverse stages.

For an adiabatic reactor the appropriate system of mass and heat transfer reads

$$\frac{\partial C_X}{\partial t} = D_X \frac{\partial^2 C_X}{\partial z^2} + \frac{j}{S}\frac{\partial C_X}{\partial z} - k_1 C_X + k_2 C_Y , \qquad (2)$$

$$\frac{\partial C_Y}{\partial t} = D_Y \frac{\partial^2 C_Y}{\partial z^2} + \frac{j}{S}\frac{\partial C_Y}{\partial z} + k_1 C_X - k_2 C_Y , \qquad (3)$$

$$\frac{\partial T}{\partial t} = \bar{\chi}\frac{\partial^2 T}{\partial z^2} + \frac{j}{S}\frac{\partial T}{\partial z} + \frac{\Delta H}{\bar{\rho}\bar{c}_p} , \qquad (4)$$

where $C_X$, $C_Y$ are the concentrations of components $X$ и $Y$; $C_{X0}$ is the inlet concentration of $X$; $D_X, D_Y$ are the diffusion coefficients of the reagents; $t, z$ are the time and space coordinates; $j$ is the integral consumption of reagents through the reactor; $T$ is the temperature; $\bar{\chi}$ is the averaged temperature conductivity; $\bar{\rho}$ is the averaged density of the reagents mixture; $\bar{c}_p$ is the averaged specific heat; $\Delta H$ is the total latent heat of the reaction; $S$ is the reactor cross-section surface.

The stationary regimes and conditions of their stability have been investigated by methods of hydrodynamic stability theory and numerical experiment. As a result, the simultaneous governing transfer parameters including thermal, diffusion and kinetic characteristics of both reagents with allowance for the two reaction stages were detailed. The existence conditions for dissipative structures which can be identified as running circular cells or wave fronts have been obtained. The stable oscillating structures, decreasing oscillations and destroying structures were considered.

In particular, it has been clearly defined a role of rate constants dependencies on temperature with allowance for direct and reverse stages of the first-order chemical reaction. The main types of possible dissipate structures induced by these factors in a non-isothermal tubular through-reactor as well as conditions of their formation have been determined. The results of investigations are likely to be useful for calculating intensity of heat and mass transfer processes in chemical apparatuses.

Using the theoretically derived conditions the minimum length of reactor in which the space dissipative structures may be realized has been obtained as function of the set of dimensionless governing parameters.

This paper deals also with a problem of gas absorption accompanied by an instantaneous, irreversible reaction in the liquid layer. The well-known methods for calculating such processes are based usually on the certain amendments to solutions, which are obtained disregarding the chemical reaction. Unlike the known work [1] the approach we used takes into account the influence of reaction resulting product on the arising and velocity of a moving reaction plane.

Increase in unit power of chemical apparatuses on conditions maintaining even distribution of the phases in a apparatus, allows reducing specific quantity of materials and energy. At the same time the maintenance of the even phase distribution on entire apparatus volume is a complex technical problem. It is especially true for large-scale apparatuses. And for uneven distribution of phases the flows ratios which are optimum from the point of view of maximum conversion, turn out to be violated in different areas of an apparatus. As a result the average conversion decreases. This phenomenon is known as large-scale effect [2]. The well-known results in the theory of chemical apparatuses scaling [3] are devoted to apparatuses with non-regular packings mainly. However how the phases distribution over the regular packings of chemical columns effects the heat and mass efficiency is studied lesser.

It is obvious that both liquid phase and gas phase distributions have significant influence on the efficiency of mass transfer in a chemical apparatus. And one of the interaction phases is usually dispersed. However, on our opinion, the dispersed phase distribution deserves of more intent consideration. Indeed, it is rather difficult to reach the even distribution of dispersed phase on entire apparatus volume. Besides, simulation of the structure of dispersed phase flow is more complex problem than the one for a continuous phase [4].

Our study was intended to the situation when the dispersed phase is liquid. This situation is characteristic for packed towers under the regimes of a moderate liquid load.

The character of liquid distribution on a packed tower depends both on the liquid distribution at the initial cross section of an apparatus, i.e. at the zone of spraying devices, and on the design of packing [5].

The evenness of liquid distribution in the apparatus cross section as a result of exposure to the packing, as the distance from the initial irrigated cross section increases, can be improved or, conversely, can become worse.

Our study was intended to the situation when the dispersed phase is liquid. This situation is characteristic for packed towers under the regimes of a moderate liquid load.

The character of liquid distribution on a packed tower depends both on the liquid distribution at the initial cross section of an apparatus, i.e. at the zone of spraying devices, and on the design of packing.

The evenness of liquid distribution in the apparatus cross section as a result of exposure to the packing, as the distance from the initial irrigated cross section increases, can be improved or, conversely, can become worse.

Our paper deals with the problem of evaluating the influence of large-scale effect on mass transfer efficiency of chemical apparatuses.

**1. Modeling the sorption process in reactor with moving reaction plane**

The real sorption processes, which are used for gas purifying or removing valuable components from gaseous mixtures, are accompanied by chemical reactions in the bulk of liquid layer and on its surface. As a rule, these reactions are multistage and, moreover, kinetic constants of different stages vary on the wide range [1, 6].

The case of two reaction stages often occurs [7]. At first, the transformation of absorbed component into the active form is going with the finite speed. Then, this form instantaneously reacts with a non-volatile active component of absorbent. As a whole the absorption proceeds in the mixed diffusion-kinetic zone.

In the course of a chemo-sorption process the concurrence of kinetic and diffusion aspects in the different zones of liquid layer leads to a very complex picture of the distribution of inlet components and reaction resulting products in the layer. As a result there are formed some reaction planes which divide the liquid layer into the parts with different ratios between diffusion and kinetic resistances. These phenomena are essentially non-stationary, and reaction planes move with time-depending velocities. Up-to-day the general approach to describing that kind of processes is absent.

Our paper deals with a problem of gas absorption accompanied by an instantaneous, irreversible reaction in the liquid layer. The well-known methods for calculating such processes are based usually on the certain amendments to solutions, which are obtained disregarding the chemical reaction. Unlike the known work [1] the approach we used takes into account the influence of reaction resulting product on the arising and velocity of a moving reaction plane.

**I. Mathematical model**

Let's consider the absorption accompanied by an instantaneous irreversible reaction:

$$A + B \to E, \qquad (5)$$

where $A$ is the absorbed gaseous component, $B$ is the active liquid-phase reactant and $E$ is the product of reaction.

The dependence of the practical diffusion coefficients on the concentrations of components is, as a rule, weak on the wide range of concentrations [8]. Thus, the fluxes of components $B$ and $E$ in the liquid phase can be written as:

$$\begin{aligned} j_B &= -D_{BB}\frac{\partial C_B}{\partial X} - D_{BE}\frac{\partial C_E}{\partial X}, \\ j_E &= -D_{EB}\frac{\partial C_B}{\partial X} - D_{EE}\frac{\partial C_E}{\partial X}, \end{aligned} \qquad (6)$$

where $D$ are the practical diffusion coefficients; $C$ denotes the concentrations of components; $X$ is the axis which directs into the layer depth normal to liquid layer surface. The indices are clear without comments.

On the initial period $t<t^*$ the reaction plane coincides with the liquid surface ($X=0$). Thus in the liquid depths the mass sources are absent. Therefore, we can write the conservation laws in the form:

$$\frac{\partial C_B}{\partial t} + \nabla j_B = 0,$$
$$\frac{\partial C_E}{\partial t} + \nabla j_E = 0. \tag{7}$$

From this it follows the diffusion equations for components $B$ and $E$ at $X>0$, $t>0$:

$$D_{BB}\frac{\partial^2 C_B}{\partial^2 X} + D_{BE}\frac{\partial^2 C_E}{\partial^2 X} = \frac{\partial C_B}{\partial t},$$
$$D_{EB}\frac{\partial^2 C_B}{\partial^2 X} + D_{EE}\frac{\partial^2 C_E}{\partial^2 X} = \frac{\partial C_E}{\partial t}. \tag{8}$$

The reaction plane becomes mobile and starts to move deep into the liquid layer at the moment $t^*$ when $B$ concentration on the liquid surface becomes equal to zero.

The diffusion equations for components $A$ и $B$ at $X>0$, $t>t^*$ read

$$D_{AA}\frac{\partial^2 C_A}{\partial^2 X} + D_{AE}\frac{\partial^2 C_E}{\partial^2 X} = \frac{\partial C_A}{\partial t},$$
$$D_{EA}\frac{\partial^2 C_A}{\partial^2 X} + D_{EE}\frac{\partial^2 C_E}{\partial^2 X} = \frac{\partial C_E}{\partial t}. \tag{9}$$

Accounting the reaction equation and expressions for fluxes the boundary conditions for $B$ and $E$ at $X=0$ look as follows:

$$D_{BB}\frac{\partial C_B}{\partial X} + D_{BE}\frac{\partial C_E}{\partial X} = \alpha C_{A\infty},$$
$$D_{EB}\frac{\partial C_B}{\partial X} + D_{EE}\frac{\partial C_E}{\partial X} = -2\alpha C_{A\infty}, \quad t>0 \tag{10}$$

$$\frac{\partial C_E}{dX} = 0, \; t>t^* \tag{11}$$

where $\alpha$ is the mass transfer coefficient into the gas phase; $C_{A\infty}$ is the concentration of $A$ into the gas phase core. Expressions (8), (9) correspond to the time when $A$ reacts with $B$ only on the liquid surface. So, component $A$ during this time can't penetrate into the liquid layer. Therefore, the equilibrium concentration $C^*_A$ is equal to zero. Component $E$ is also supposed to be non-volatile.

Boundary conditions for component $E$ at $X=\infty$, $t>0$:

$$C_E=0, \tag{12}$$

for component $B$ at $X=\infty$, $t>0$

$$C_B=C_{B\infty}, \tag{13}$$

Boundary conditions for components $A$ и $B$ at $X=y(t)$, when $t>t^*$

$$C_A=C_B=0,\ x=y(t),\ t>t^*. \tag{14}$$

Relation between $A$, $B$ and $E$ fluxes read:

$$j_A + j_B = j_E = j_{E_1} + j_{E_2} = 2j_A = 2j_B \tag{15}$$

Appropriate boundary conditions are

$$2j_B = j_E, \tag{16}$$

$$j_A = j_B \tag{17}$$

From (16), (17) it follows (Fig. 1, I zone)

$$2\left(D_{BB}\frac{\partial C_B}{\partial X} + D_{BE}\frac{\partial C_E}{\partial X}\right) = \left(D_{EE}\frac{\partial C_E}{\partial X} + D_{EA}\frac{\partial C_A}{\partial X} + D_{EE}\frac{\partial C_E}{\partial X} + D_{EB}\frac{\partial C_B}{\partial X}\right) \tag{18}$$

$$D_{BB}\frac{\partial C_B}{\partial X} + D_{BE}\frac{\partial C_E}{\partial X} = -\left(D_{AA}\frac{\partial C_A}{\partial X} + D_{AE}\frac{\partial C_E}{\partial X}\right), \tag{19}$$

$$D_{EE}\frac{\partial C_E}{\partial X} + D_{EA}\frac{\partial C_A}{\partial X} = 2\alpha C_{A\infty}, \tag{20}$$

The appropriate expression for the zone II reads:

$$D_{EE}\frac{\partial C_E}{\partial X} + D_{EB}\frac{\partial C_B}{\partial X} = -2\alpha C_{A\infty}, \tag{21}$$

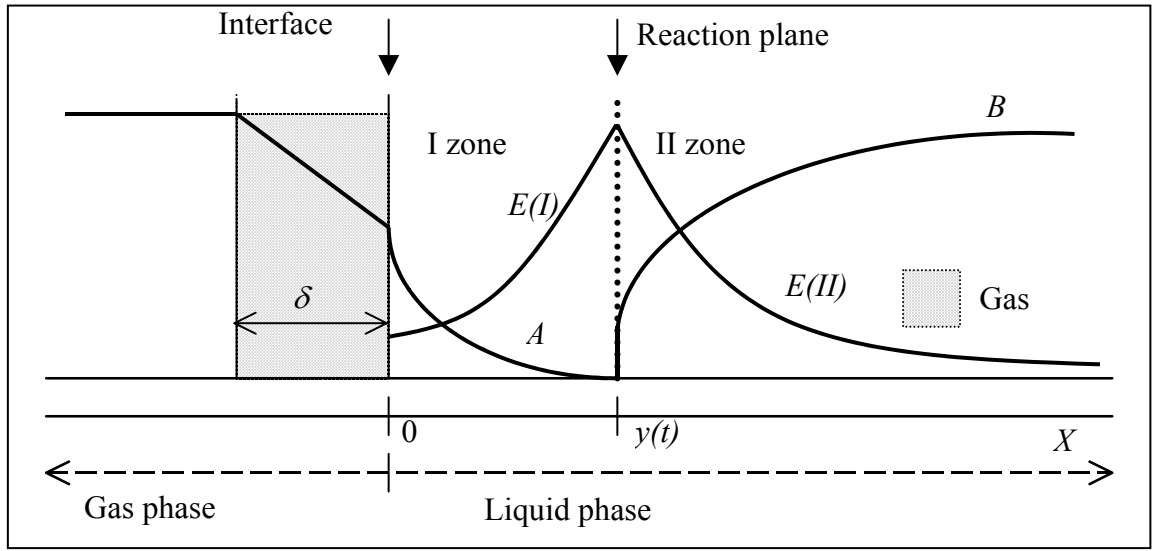

Figure 1: Profiles of components concentrations ($t > t^*$)

The system (5) with boundary conditions (12) - (17) can be solved by means of Laplace transformation. The Laplace images of concentrations are:

$$\tilde{C}_B = \frac{\alpha C_{A\infty}}{p\sqrt{p}(R_2 - R_1)}\left[\frac{2T_2 + T_1}{\sqrt{S_1}}\exp(-\sqrt{S_1 p}X) - \frac{2T_2 R_1 + T_1 R_2}{\sqrt{S_2}}\exp(-\sqrt{S_2 p}X)\right] + \frac{C_{B\infty}}{p}, \quad (22)$$

$$\tilde{C}_E = \frac{\alpha C_{A\infty}}{p\sqrt{p}(R_2 - R_1)}\left[\frac{2T_2 + T_1}{\sqrt{S_1}}\lambda_1\exp(-\sqrt{S_1 p}X) - \frac{2T_2 R_1 + T_1 R_2}{\sqrt{S_2}}\lambda_2\exp(-\sqrt{S_2 p}X)\right] \quad (23)$$

As a result, we get the following expressions for the concentrations $B$ and $E$ into the liquid layer for the time preceding the moment $t^*$.

$$\tilde{C}_B = \frac{\alpha C_{A\infty}}{(R_2 - R_1)}\frac{2T_2 + T_1}{\sqrt{S_1}}\left[2\sqrt{\frac{t}{\pi}}\exp\left(-\frac{x^2 S_1}{4t}\right) - x\,\text{erfc}\left(\frac{x\sqrt{S_1}}{2\sqrt{t}}\right)\right] - \frac{\alpha C_{A\infty}}{(R_2 - R_1)}\frac{2T_2 R_1 + T_1 R_2}{\sqrt{S_2}}\left[2\sqrt{\frac{t}{\pi}}\exp\left(-\frac{x^2 S_2}{4t}\right) - x\,\text{erfc}\left(\frac{x\sqrt{S_2}}{2\sqrt{t}}\right)\right] + C_{B\infty}, \quad (24)$$

$$\tilde{C}_E = \frac{\alpha C_{A\infty}}{(R_2 - R_1)}\frac{2T_2 + T_1}{\sqrt{S_1}}\lambda_1\left[2\sqrt{\frac{t}{\pi}}\exp\left(-\frac{x^2 S_1}{4t}\right) - x\,\text{erfc}\left(\frac{x\sqrt{S_1}}{2\sqrt{t}}\right)\right] - \frac{\alpha C_{A\infty}}{(R_2 - R_1)}\frac{2T_2 R_1 + T_1 R_2}{\sqrt{S_2}}\lambda_2\left[2\sqrt{\frac{t}{\pi}}\exp\left(-\frac{x^2 S_2}{4t}\right) - x\,\text{erfc}\left(\frac{x\sqrt{S_2}}{2\sqrt{t}}\right)\right] \quad (25)$$

The appropriate concentrations on the liquid surface are:

$$C_{BS} = \frac{2\alpha C_{A\infty}}{(R_2 - R_1)}\left(\frac{2T_2 + T_1}{\sqrt{S_1}} - \frac{2T_2 R_1 + T_1 R_2}{\sqrt{S_2}}\right)\sqrt{\frac{t}{\pi}} + C_{B\infty}, \quad (26)$$

$$C_{ES} = \frac{2\alpha C_{A\infty}}{(R_2 - R_1)}\left(\lambda_1\frac{2T_2 + T_1}{\sqrt{S_1}} - \lambda_2\frac{2T_2 R_1 + T_1 R_2}{\sqrt{S_2}}\right)\sqrt{\frac{t}{\pi}}. \quad (27)$$

Thus, from the condition $C_{BS} = 0$ we obtain:

$$t^* = \frac{C_{B\infty}^2 \pi (R_2 - R_1)^2}{4\alpha^2 C_{A\infty}^2\left(\frac{2T_2 R_1 + T_1 R_2}{\sqrt{S_2}} - \frac{2T_2 + T_1}{\sqrt{S_1}}\right)^2} \quad (28)$$

At this moment the concentration of reaction product $E$ on the interface is maximum.

$$C_{ES} = \frac{C_{B\infty}\left[\lambda_1\sqrt{S_2}(2T_2+T_1) - \lambda_2\sqrt{S_1}(2T_2 R_1 + T_1 R_2)\right]}{\left[\sqrt{S_2}(2T_2+T_1) - \sqrt{S_1}(2T_2 R_1 + T_1 R_2)\right]} \quad (29)$$

## II. Numerical experiments

For the time $t>t^*$ systems (8),(9) can't be solved analytically [1]. Therefore, we used Crank-Nicholson method modified in comparison with [1].

Let's consider the lattice (Fig.2, Fig.3) on the time levels: $t_l$, $t_{l+1} = t_l + \Delta t_l$. Let $0 \equiv x_0 < x_1 < x_2 ... x_{2N} \equiv L$ be a fixed lattice in the liquid layer, where $L$ is large but finite value.

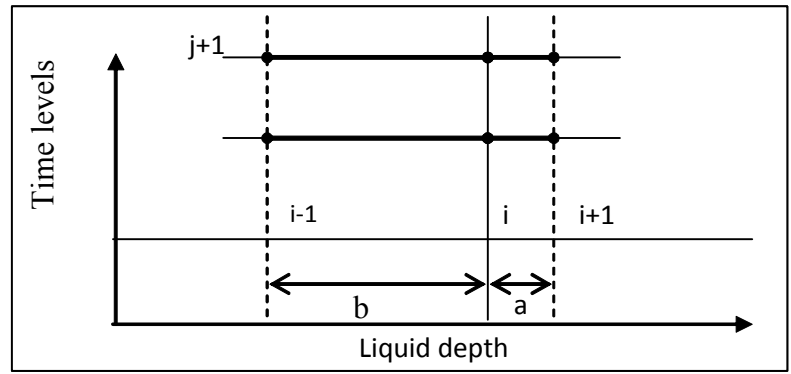

Figure 2: Lattice of points

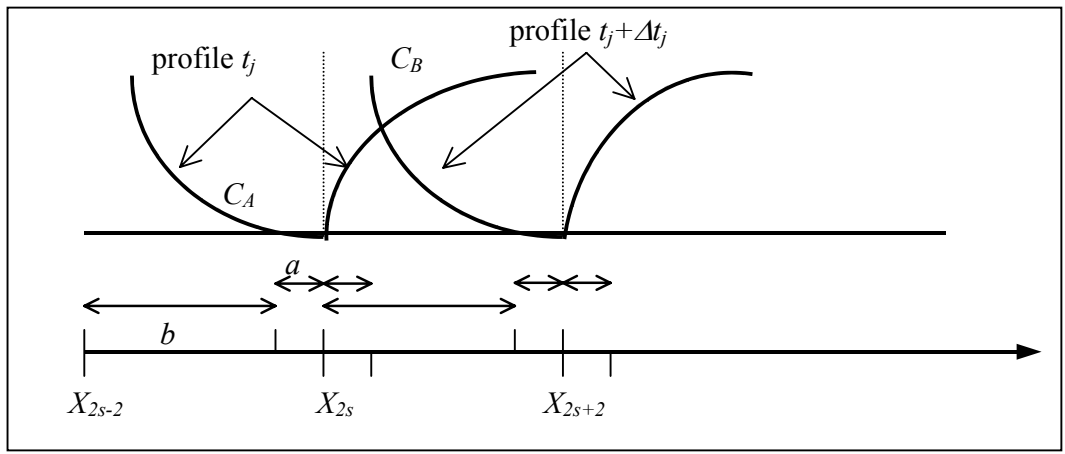

Figure 3: Concentration profiles at two subsequent time points

For numerical solving the finite differential equations system we used the method of separate driving, which is illustrated in Fig. 4. The gist of this method is following. The finite differential equations for the zone II are divided into two groups. Group 1 corresponds to the equations for the concentration $B$, and group 2- for the concentration $E$. Equations of group 1 are solved by means of driving method. Concentration of the component $E$ during this driving process doesn't change. The value of $E$ is taken from the preceding level $j$.

When finishing the calculations in the group 1, the obtained value of concentration $B$ transfers to the group 2 for calculating the value of $E$ by means of the driving method. The last value is used in the exterior iteration on the level $(j+1)$ for group 2. Then, the obtained values of $B$ и $E$ are used on the level $(j+2)$ etc.

Note that the main features of the method coincide with the ones used in [1].

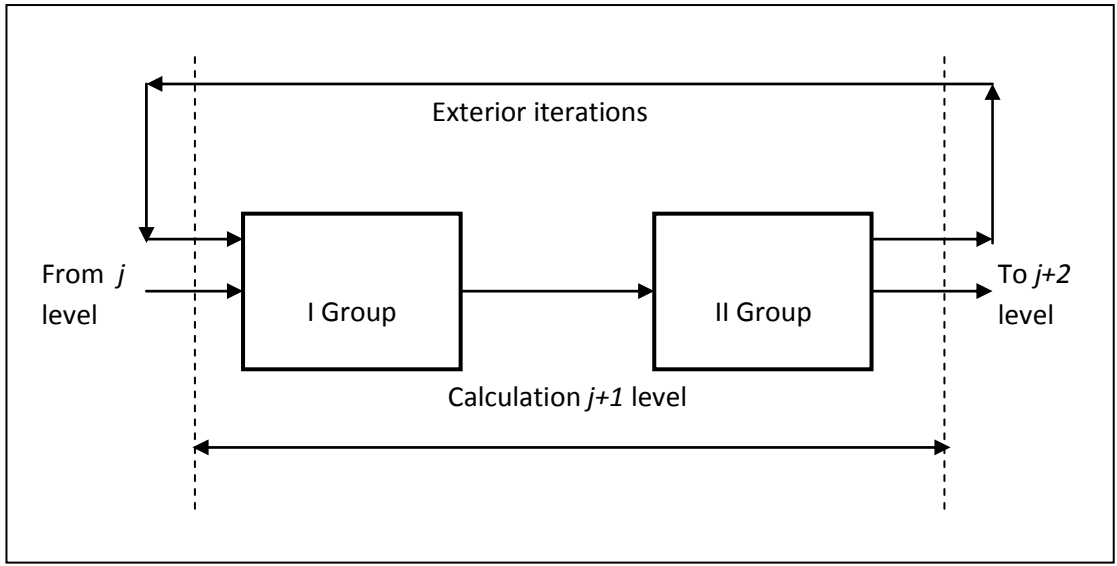

Figure 4: Scheme of the separate driving method

The only but important peculiarity is that while calculating the value of $E$ at the reaction plane we can obtain disparity between values of $E1$ and $E2$ in the nearest points on the left and on the right from the reaction plane. Therefore, we needed special iterations for coordinating the calculated concentrations of component $E$ on the reaction plane and fluxes of components on both sides of the plane.

On the base of the mathematical model the series of numerical experiments has been carried out. We investigated in details the influence of the set of governing parameters: partial pressure of gaseous component $A$; concentration of the active liquid- phase reactant $B$; main and cross diffusion coefficients; Henry's constant $H$ and mass transfer coefficient in gas phase $\alpha$ on the reaction plane velocity; concentration of the reaction product $E$ and concentration of $A$ - component on the liquid surface.

The most principal result of our numerical experiments lies in the finding of the some short period $(t^* < t < t_p)$ when the velocity of reaction plane quickly increases and reaches a maximum. Then, after this period $(t > t_p)$, the reaction plane velocity monotonously decreases and slowly approaches to zero when time becomes infinite.

The two characteristic time sites are clear seen in Fig. 5. On the base of our results we can estimate the duration of the first period $t^* < t < t_p$ of reaction plane movement and location of this plane $h_p$ at the moment $t = t_p$. Note that such estimations can't be obtained from other known models. The second period $t > t_p$ is well-known from the literature and described by the film model [4]. The knowledge of the characteristic time $t_p$ and the characteristic penetration depth $h_p$ is very important for defining the bounds of film model correctness.

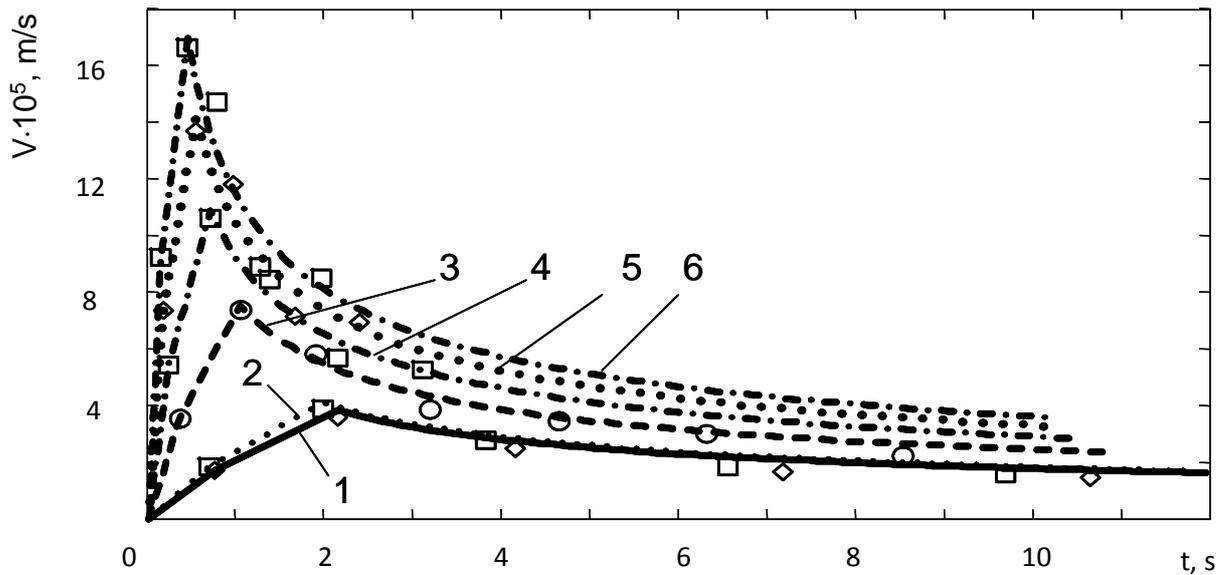

1 - $D_A$=0.73·10⁻⁹, 2 - $D_A$=1.73·10⁻⁹, 3 - $D_A$=2.73·10⁻⁹, 4 - $D_A$=3.73·10⁻⁹,
5 - $D_A$=4.73·10⁻⁹, 6 - $D_A$=5.73·10⁻⁹ m²/s
Figure 5: Time-dependent velocity of the reaction plane

It was established from numerical experiments that increase of the partial pressure of $A$-component leads to increasing the maximum velocity of reaction plane $V_f$. By increasing the $B$-concentration in the liquid layer core the maximum of $V_f$ also increases.

When increasing the main diffusion coefficients the concentration of product $E$ on the reaction plane and penetration time $t_p$ essentially decrease.

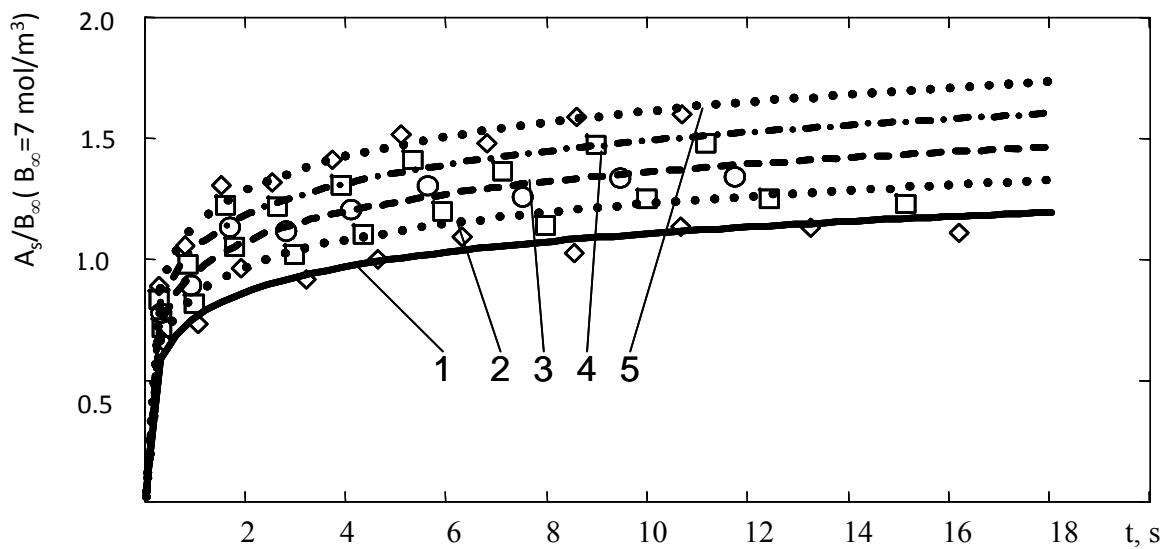

1- $p_A$=0.9·10³ Pa, 2 - $p_A$=1·10³ Pa, 3- $p_A$=1.1·10³ Pa, 4 - $p_A$=1.2·10³ Pa, 5 - $p_A$=1.3·10³ Pa
Figure 6: Time-dependent concentration of the absorbed component $A$ on the liquid surface

It was shown that in the first penetration period $t^* < t < t_p$ the concentration of $A$-component on the liquid layer surface quickly increases, but after $t_p$ this concentration is practically constant. This result agrees with known experimental data [1, 8].

Fig.7 depicts the time history of the surface concentration of reaction product $E$. From this figure we can see that $E_{fr}$ monotonously decreases during the time.

After processing the data of numerical experiment we were able to obtain the following generalized formulae for calculating the main characteristics of absorption (period $t^* < t < t_p$).

1. Characteristic penetration time, s:

$$t_p = 13.6\left(1 + 0.018\frac{C_B H}{P}\right)^{0.786}\left(\frac{D_A}{D_B}\right)^{-3.44}. \tag{30}$$

2. Characteristic penetration depth:

$$h_p = 7.72\cdot 10^{-5}\left(1 + \frac{D_E}{D_B}\right)^{-0.18}\left(\frac{D_A}{D_B}\right)^{0.781}. \tag{31}$$

3. Concentration of the absorbed component on the interface:

$$\frac{C_{AS}}{C_B} = \left(2.371 - \frac{C_B H}{P}\right)^{0.802}\left(\frac{t}{t_p}\right)^{0.408} \tag{32}$$

4. Concentration of reaction product on the interface:

$$\frac{C_{Efr}}{C_{Efr}(0)} = \exp\left[-\left(\frac{D_E}{D_A}\right)^{0.45}\left(\frac{t}{t_p}\right)^{0.652}\right]. \tag{33}$$

5. Averaged mass transfer coefficient in the liquid phase:

$$\overline{\alpha_l} = \alpha\left[2.451\frac{P}{C_B}\left(2.371 - \frac{C_B H}{P}\right)^{-0.802} - H\right] \tag{34}$$

Using these formulae we can determine the acceleration coefficients of mass transfer [8] and adapt our method to engineering calculations.

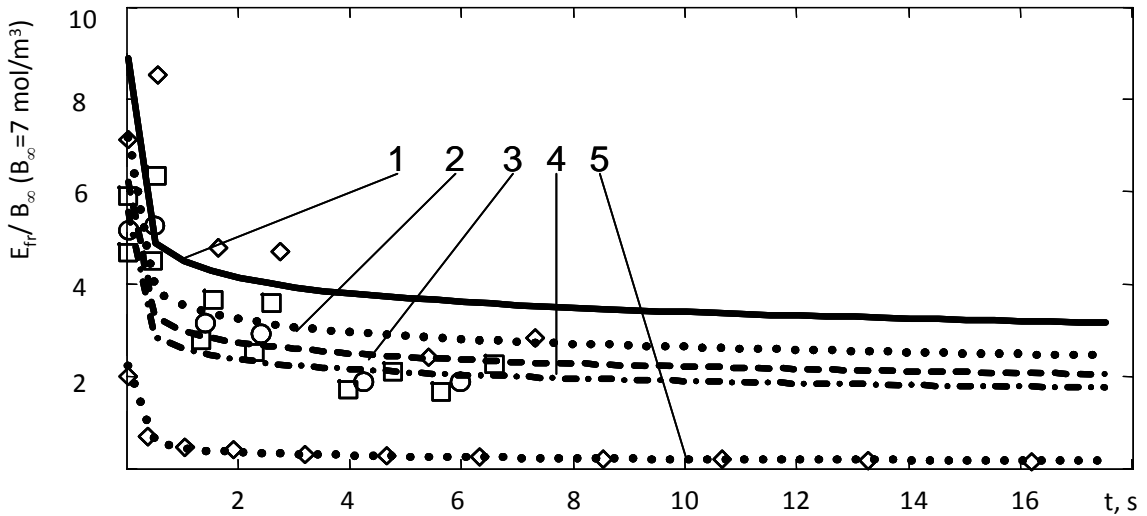

1- $D_E=0.04\cdot 10^{-9}$, 2 - $D_E=0.06\cdot 10^{-9}$, 3 - $D_E=0.08\cdot 10^{-9}$, 4 - $D_E=0.1\cdot 10^{-9}$, 5 - $D_E=2.1\cdot 10^{-9}$ m$^2$/s
Figure 7: Time-dependent concentration of the reaction product $E$ on the liquid surface

### III. Mixing process

The other important problem is linked with the peculiarities of mixing process in the presence of chemical conversion.

The key parameters for the mathematical modeling of fluxes structure in through-reactors as well as for describing conditions of reactants conversion are the three time scales: the time of the element life in the reactor work volume $\bar{t}$; the age of the element $\alpha$, i.e. the time from inlet moment to the moment of the observation and the expecting time of the element $\lambda$, i.e. the period between the observation moment and the moment of outlet from a system.

The two extreme states of the micro-mixing have been considered. They are the complete segregation and the complete mixing up. The first state means that the every element remains in the vicinity of elements with the same ages all the time while staying in the system.

By this approach the elements distribution in the system is described according their ages. The second extreme state means instant mixing up of all elements with the same expecting times.

The rate of the transfer process from the state of complete segregation to the state of complete micro-mixing up reads

$$dG = \gamma V d\lambda \frac{1}{\bar{t}} \int_0^\infty e^{-\gamma \alpha} F(\alpha + \lambda) d\alpha, \qquad (35)$$

where $V$ is the reactor volume; $\bar{t}$ is the average life time of elements in the reactor; $\gamma$ is the specific rate parameter.

Mass balance of for differential volume from reagents $\lambda$ to $\lambda+d\lambda$ at non-stationary conditions for the i-th component while substance transition from the segregate state to the micro-mixing medium looks as follows

$$dG_2 C_i[(\lambda + d\lambda), t]dt - dG_2 C_i(\lambda, t)dt + dG_3(C_{exi} - C_{cp})\gamma d\lambda dt -$$
$$dG_2 R(C_i, t, T) d\lambda dt = \pm dG_1 dC_i(\lambda, t) d\lambda \qquad (36)$$

Where the first term is the inlet volume to the medium of an interim state, the second term is the outlet flux, the third term is the flux of substance transferred from main flow and the fourth term is the chemical mass source. The right part of the equation shows an accumulation or an expense of the substance in the indicated zone.

On base of this approach we consider the model of micro-mixing structure as sequence of connecting cells with the complete segregation and complete mixing up. The concrete structure of these cells system is depended on the mixing regime and apparatus design. The main governing parameter in this model is the crucial age of segregating elements. It means that excess of this age leads to the level of complete micro-mixing.

The results of mathematical modeling were compared with experimental data. The dynamical characteristics were investigated in the liquid-phase reactor of 0.67 m$^3$ volume.

The optimum constructive and technological parameters were obtained, and carried out methods and experimental data were used for designing the industrial reactor for reactive phosphorus salts production.

### 2. The concept of random walk methods for describing the liquid phase distribution on regular packings

Most studies on the distribution of dispersed liquid phase over the packings have been performed for the film regime of liquid flow over the surface of packing units, when the gas load has little effect on the liquid flow.

Currently, there is no general theory of liquid distribution over various packings that would adequately describe the effects of all constructive and regime parameters. This can be explained by the extreme complexity of hydrodynamic situation that is formed in the apparatus when the dispersed liquid flows through the packed bed in contra-current with the gas stream [9].

The random walk methods were successfully used for describing the liquid distribution over the regular shelf packings [10]. At the same time there is no apparent cause why these methods can not be adapted for describing liquid distribution over regular packings of various types.

The approach [11] is founded on two main assumptions.

The first is that probability for the system to change its state from the one state to other depends on these pairs of states only, i.e. the system states form the Markov chain. The assumption lies on that the intensity of liquid stream on the packing unit depends on local space orientation of this unit and certain neighboring units only, but the structure of local liquid flow is supposed to be independent on local stream intensity. It may be correct to some extent.

The second assumption is that increase of distance between two packing units leads to rapid decrease of the probability for the liquid direct overflowing from one of the mentioned packing units to another. For many types of packings this assumption will be practically fulfilled provided sizable non-uniformity of the packing is absent [12].

Whether the packing design is that allows formation of more complex structure of liquid flow over the packing units then we can formulate the uniformity condition as the probability of liquid overflowing depends only on the difference $|x_i - x_j|$. Thus, it is possible to give the adequate mathematical description of the distribution of the dispersed phase on volume of the packing tower, even with complicated internal devices, with the help of stochastic methods, in particular, the random walk methods. Therefore we can talk about the prospects of the development of such methods to describing the structure of the flow phases in chemical apparatus.

In the works [13] the methods of random walk were used for the mathematical description of liquid distribution over the regular packing. The results of computer simulation that was carried out for regular chord and shelf packings agreed well with the experimental data in the case when packing is watered with a system of point sources of irrigation.

Thus, the local specific liquid flow intensity $i_{rz}$ while packing into the column of diameter $D$ watered by one axis-symmetric spray is determined by the following formula [14]

$$i_{rz} \approx I \frac{2h}{\pi z} \left\{ \exp\left(-\frac{h}{2a^2 z} r^2\right) + \exp\left(-\frac{h}{2a^2 z}(D-r)^2\right) + \exp\left(-\frac{h}{2a^2 z}(D+r)^2\right) \right\} \qquad (37)$$

In the presence of set of point sources of irrigation the local flow intensities can be evaluated with the help of the methods of local sources superposition [15].

In normal case, under the even distribution of liquid over the regular packing the liquid flows down from each packing unit to the certain number of downstream units. This number depends on the design both of the packing units and of the packing as whole. Studies show that under the certain gas velocities and liquid flow intensity that are optimum for design of given packing there exist conditions for establishing the uniform liquid distribution over the cross section of an apparatus at some distance from the initial section where the liquid sprayers have set.

The size of the spreading zone $H_s$ can be defined as the distance from the cross-section irrigated to the cross-section, wherein the unevenness distribution coefficient, i.e. the ratio of the maximum of the local liquid flow intensity to the minimum of the local intensity reaches a predetermined optimum value. As it is shown by calculations and confirmed by experimental data, the irrigation intensity maxima are observed under the liquid sources up to the certain distance from initial cross-section.

As the distance from the irrigated section increases the maximum of the local liquid intensity, in the absence of bypass in the peripheral zone, is formed on the axis of the apparatus. In the presence of draining nearby the walls the maximum of intensity may be formed at the walls of the apparatus on some distance from the initial section. Such hydrodynamic regime leads to a significant reduction in process efficiency [16].

Thus, the full height of the apparatus can be divided into two zones. In the first zone, which can be designated as zone of liquid spreading, the liquid distribution over the cross section of the column may differ significantly from even distribution. In the second zone the distribution is close to uniform in the absence of obvious defects of the packing units or the packing assembly. In order to determine the characteristics of absorption with account of the liquid distribution it is necessary to know the dependence of the specific mass transfer coefficients on the local intensity of irrigation. Such approach allows to simplify the mathematical model and to propose also a simplified technique of engineering calculations.

Let us define the coefficient of the distribution unevenness $k_u$ at the end of the spreading zone as

$$k_u = \frac{i_{0,H_s}}{i_{R,H_s}} \qquad (38)$$

Geometric properties of the irrigation device can be characterized with the help of the parameter $n$ - a number of point sources of irrigation per unit of cross-section. Geometric parameters of packing are a typical hydrodynamic radius of the packing unit $a$ and a typical height of the layer of packing $h$.

Figure 8 depicts some results of the numerical experiment that have been carried out according to random walk model [2].

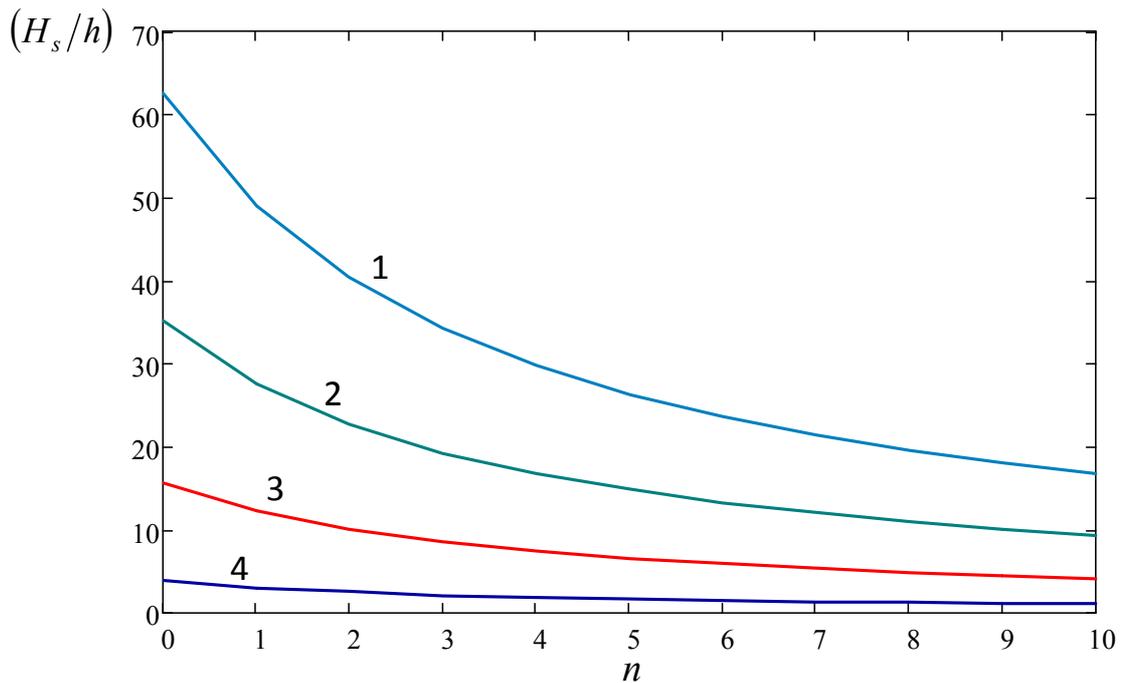

*Figure 8: Dimensionless height of the spreading zone $(H_s/h)$ as function of the number $n$ of point sources of irrigation per unit of cross-section for chord packing. $D/a$ = 20(1); 15(2); 10(3); 5(4), $k_u = 1.15$*

On the results of numerical experiment the following approximation was derived [17]

$$\frac{H_s}{h} = \frac{(D/a)^2}{4.64 + 1.76n} \qquad (39)$$

Analysis of the results of numerical experiments shows that increasing of the number of sources of irrigation can save or decrease the height of spreading zone only under increasing $n$, what will be fulfilled when step between sources at the irrigating cross-section decreases.

Another principal result of the analysis of liquid distribution model is that there exists characteristic radius $R_a$ for which the local intensity of liquid flow is equal to average value calculated for complete cross-section. Moreover, this characteristic radius becomes constant at certain distance $H_a$ from the initial cross-section, i.e. $R_a$ does not change further over the apparatus height.

The following estimations for the radius $R_a$ and corresponding to it average intensity of the liquid flow has been obtained

$$R_S = \sqrt{\frac{aD}{2}\ln\left(\frac{4D}{\pi a}\right)} \tag{40}$$

$$\bar{j} = J\sqrt{\frac{2h}{H_S}}\exp\left(-\frac{hR_S^2}{2a^2 H_S}\right) \tag{41}$$

### 3. Evaluation of large-scale effect from the point of liquid distribution

The influence of uneven liquid distribution on the efficiency of absorption in the spreading zone is determined by three main reasons. They are: decrease of the specific surface of liquid dispersion due to low intensity of irrigation at the part of packing units, dependence of the local mass transfer coefficients on the local liquid flow intensity and, finally, varying absorption factor [18] over the apparatus cross-section. Of course, these reasons are interconnected, and acting in common they lead to decreasing the average mass transfer coefficient into the spreading zone as compared with beyond of the spreading zone where more or less uniform liquid distribution can be established. In the works [3] the reduction of the mass transfer coefficient due to uneven liquid distribution has been described with the help of so called "coefficient of worsening".

In our work we offer a semi-empirical approach according to which the coefficient of worsening is a ratio

$$\gamma = \frac{k_{ms}}{k_m} \tag{42}$$

where $k_{ms}$ is the average mass transfer coefficient into the spreading zone; $k_m$ is the empirical mass transfer coefficient corresponding to the average absorption factor outside the spreading zone.

Thus two cases should be distinguished. In the first case, the average intensity of irrigation is sufficient to achieve the optimal regime of intensive mass transfer. In the second case, the average intensity of irrigation may not be sufficient to achieve the optimum performance for mass transfer.

Let us further assume that the first case is realized. If the main resistance to mass transfer is concentrated in the gas phase that of all influencing factors the reduction of the specific surface of contact between the phases is brought to the forefront.

The influence of liquid distribution on the mass transfer intensity and efficiency of absorption process in the column with regular shelf packing was investigated on the pilot plan [19]. Known experimental results confirm this thesis (Figure 9).

While analyzing the experimental results it can be made some conclusions. First, growth of the average intensity of the irrigation leads to increase of the coefficient of worsening, and it becomes stable at a certain level depending on the number of point sources of irrigation. Such behavior of the coefficient of worsening completely corresponds to the theoretical conceptions described above.

Second, the local values of the mass transfer coefficients are got from empirical data. It means that for accounting the uneven liquid flow distribution it is assumed that each point of the apparatus volume element can be conventionally associated with the local value of the mass transfer coefficient obtained in the laboratory facilities of small size with a known structure of interacting flows [20].

Thus, under the rather high specific intensity of irrigation the coefficient of worsening can be calculated from mere geometrical considerations, namely – as a ratio between a volume of a watered part of the spreading zone and its complete volume. For a local minimum of liquid flow rate on the basis of stochastic random walk model we obtain:

$$i_{min} = \frac{4I\sqrt{\frac{a}{\pi d_0}}}{\exp\left(-\frac{d_0}{4a}\right)} \qquad (43)$$

Here, the geometrical parameter of the initial liquid distribution characterized by some conditional step $d_0$ between point sources of irrigation.

On the base of the distribution model the expressions for calculating a degree $\chi$ of substances conversion at the outlet of apparatus with allowing for the presence of two zones marked above have been obtained

$$\chi = \frac{\exp\left(\frac{\lambda-1}{G}F\overline{K}[H-(1-\gamma)]H_S\right)-1}{\lambda\exp\left(\frac{\lambda-1}{G}F\overline{K}[H-(1-\gamma)]H_S\right)-1} \qquad (44)$$

In (44) $F$ is the total cross section surface, $G$ is the flow rate of the continuous phase.

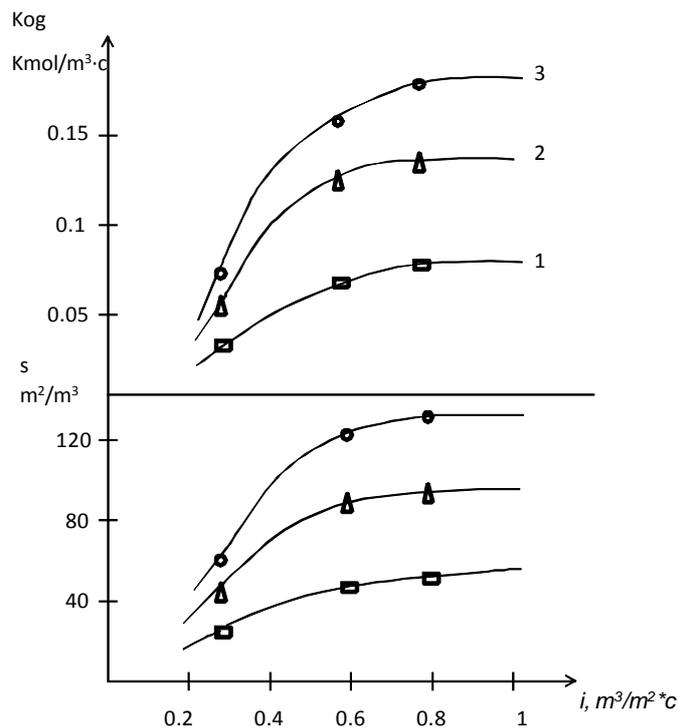

*Figure 9: Dependence of the specific interphase surface s and volume mass-transfer coefficient in the gas phase Kog on the average intensity of irrigation i: 1-$w_g$=1 m/s, 2-$w_g$=1.25 m/s, 3-$w_g$=1.5 m/s*

When evaluating the effectiveness of the device due to the uneven flow distribution in the amount of using the representation of the height of the transfer unit $\tilde{h}$ [21], the calculation of the relevant characteristics can be produced by the formulas:

$$\widetilde{h} = \widetilde{h}^* + \Delta\widetilde{h} \qquad (45)$$

$$\Delta\widetilde{h} = \frac{(1-\gamma)H_S}{N} \qquad (46)$$

$$N = \frac{1}{\lambda-1}\ln\left(\frac{1-\chi}{1-\lambda\chi}\right) \qquad (47)$$

where $\widetilde{h}^*$ is the height of the transfer unit with an uniform distribution of flows (according to experimental studies on laboratory bench), $\lambda$ is the absorption factor.

**Conclusions**

The new approach to describing the scaling effect for chemical towers with regular packings has been submitted. This approach founds on the concepts of random walk along mathematical grids applying to the dispersed phase distribution over packing with allowing for the peculiarities of phases distribution on the initial site at the vicinity of the spraying device. It is established that there exists some fixed characteristic radius, on which the flow rate of the dispersed phase (liquid) is equal to the average value over the cross section of the apparatus independent on the distance from sprayer. This radius is stabilized at a certain distance from the inlet cross section. This approach allows us to decompose the apparatus work volume onto zones with different heat and mass transfer efficiency. So the sufficiently simple methods for evaluating the influence of large-scale factor on the efficiency of mass transfer have been obtained. These methods are suitable for use in engineering calculation techniques.